\documentclass[pre,showpacs,noshowkeys,amsmath,superscriptaddress,twocolumn,letterpaper]{revtex4}
\usepackage{ae,aecompl}
\usepackage{longtable}
\usepackage{graphicx}
\begin{document}
\newcommand{\ICC}[0]{ICC$^\star$}
\newcommand{\icc}[0]{ICC$^\star$}
\newcommand{\NP}[0]{\ensuremath{N_{\mathrm pore}}}
\renewcommand{\vec}[1]{\boldsymbol{ #1}}
\renewcommand{\u}[1]{\vec #1}
\newcommand{\bra}[1]{\langle #1\rvert}
\newcommand{\ket}[1]{\lvert #1 \rangle}
\newcommand{\expon}[1]{e^{#1}}
\renewcommand{\d}{\mathrm d}

\title{Influence of pore dielectric boundaries on the  translocation barrier of DNA }
\author{Stefan Kesselheim}
\email{stefan.kesselheim@icp.uni-stuttgart.de}
\affiliation{Institute of Computational Physics, University of Stuttgart, Pfaffenwaldring 27, D-70569 Stuttgart, Germany}
\author{Marcello Sega}
\email{sega@science.unitn.it}
\affiliation{Department of Physics and INFN, University of Trento, via Sommarive 14, I-38123 Trento, Italy}
\author{Christian Holm}
\email{holm@icp.uni-stuttgart.de}
\affiliation{Institute of Computational Physics, University of Stuttgart, Pfaffenwaldring 27, D-70569 Stuttgart, Germany}

\date{\today}
\begin{abstract}
  We investigate the impact of dielectric boundary forces on the
  translocation process of charged rigid DNA segments through solid
  neutral nanopores.  We assess
  the electrostatic contribution to the translocation free energy
  barrier of a model DNA segment by evaluating the potential of mean
  force in absence and presence of polarization effects by means of
  coarse-grained molecular dynamics simulations.  The effect of
  induced polarization charges has been taken into account by
  employing \ICC, a recently developed algorithm that can efficiently
  compute induced polarization charges induced on suitably discretized
  dielectric boundaries.  Since water has a higher dielectric constant
  than the pore walls, polarization effects repel charged objects in
  the vicinity of the interface, with the effect of significantly
  increasing the free energy barrier. Another investigated side effect
  is the change of the
  counterion distribution around the charged polymer in presence of
  the induced pore charges. Furthermore we investigate the influence
  of adding salt to the solution.

\end{abstract}
\maketitle
\noindent 
\section{Introduction}
The problem of polymer translocation through nanometer-sized pores
has recently stimulated much
experimental \cite{storm05X,vanDorp09X,fologea05X,smeets06a,keyser06X,trepagnier07X,gershow07X,ghosal07X,maglia08X},
theoretical \cite{DiMarzio97X,muthukumar99X,lubensky99X,chuang01X,sakaue07X,dubbeldam07X,kotsev07X,abdolvahab08X,chatelain08X,mccauley09X}
and simulation based
\cite{milchev04X,aksimentiev04X,matysiak06X,luo06X,fyta06X,luan08X,melchionna09X}
research, due to their major role in biological processes and to
potential technological applications. Molecular transport is indeed
one of the key functions fulfilled by the plasma and nuclear membranes
of the cell, and a sizable amount of transport mechanisms
which work in the cell are characterized by the same general design:
namely, by the presence of pores, mostly through membrane proteins.  The controlled
transport of single molecules through synthetic or biological
nanopores is considered as a versatile tool of single molecule sensing
and to be a most promising candidate for rapid DNA sequencing. A
recent review reported more than 100 experimental approaches
\cite{howorka09a} that try to unveil how these systems can be
technologically used as a probe to the world of single
molecules. Researchers employed biological nanopores, or crafted
synthetic ones\cite{storm03X}, for different technological and
scientific purposes, often with an aim towards DNA sequencing
\cite{kasianowicz96a,Henrickson00X,Lagerqvist06X}.

The complex interplay of interactions -- electrostatic, hydrodynamic
and specific chemical ones -- and the entropic properties of chain
molecules makes a full understanding of these systems very
difficult. In this article we investigate a contribution to the
translocation free energy barrier of stiff DNA which has often been
neglected so far, namely the role of the dielectric mismatch between
solvent and pore.  The presence of the interface between the highly
polarizable aqueous solution ($\varepsilon \approx 80$) and the
membrane which is much less polarizable ($\varepsilon \approx 2$)
leads to repelling forces between charged objects and the pore
wall. As DNA is a highly charged molecule this effect is likely not to
be negligible and potentially gives rise to an energetic barrier that
opposes transversing the pore.  Its characteristics and dependence on
the pore size, DNA length, or salt concentration are not known. In
this work we answer some of the these questions by investigating the
translocation properties of a model, rigid DNA fragment.
Coarse-grained Molecular Dynamics (MD) simulations are employed to
compute the mean force acting on the DNA fragment, taking explicitly
into account the combined effect of the DNA counterions, salt ions at
different ionic strengths, and of surface polarization charges
generated by the presence of the dielectric mismatch.

This paper is organized as follows: in Sec.~\ref{sec:theory}, we
briefly review some aspects of theoretical models and simulation approaches for
the description of DNA translocation through pores, as well as the
strategy of our approach; in Sec.~\ref{sec:methods} we describe some
methodological details, starting with the recently developed \ICC\
(induced charge computation) algorithm \cite{tyagi10a} that accurately
calculates the induced polarization charges on the dielectric
boundaries of the pore, continuing with the description of the DNA
model and the procedure employed to estimate the free energy
profile. In Sec.~\ref{sec:results} the results of our investigation
are presented and discussed. The paper ends with some concluding
remarks in Sec.~\ref{sec:conclusions}.

\section{DNA translocation free energy\label{sec:theory}}
In order to cross a pore, a polymer undergoes several distinct phases.
The first phase is the diffusive transport of the polymer to within the vicinity of the pore entrance.
It is diffusive also under an applied external field since the applied
voltage drops only in vicinity of the pore. 

Another phase can be attributed to the polymer entering the pore which
implies a conformational change of the polymer, namely a stifffening,
that is connected to an entropic loss of the chain conformation. This
also implies that the DNA molecule experiences an interaction with the
pore along the stretched end.  This possibly imposes an additional
barrier, depending on the electrostatic and chemical interactions
within the pore.

The actual translocation process  involves the threading of the polymer through
the pore until the polymer reaches the other side of the pore. In all common experiments
the process is driven by an external electrostatic field for two reasons: The first reason
is that the ionic current through the pore is the observable that
serves for sensing. The other reason is the entrance barrier that limits the frequency with
which the polymer performs the actual translocation. This barrier is reduced
by applying an external field and thus the translocation frequency is increased.

The translocation phase turns out to be very complex for flexible
polymers. One of the major issues encountered in the description of
flexible polymers is the fact that the relaxation time of very long polymers to thermodynamic
equilibrium is longer than the translocation time
itself~\cite{chuang01X,gauthier08X}. Also when the polymer is shorter the
relaxation of the chain still appears to be strongly coupled to the translocation
process~\cite{gauthier09X}. 
This means that it is not possible to use the number of translocated monomers --- as well
as any other single generalized variable ---  as a reaction coordinate and to assume all other
degrees of freedom to be relaxed (as is the case of Muthukumar's model~\cite{muthukumar99X}).
This has a profound implication, because it shows that the translocation process is genuinely
irreversible: theoretical approaches based on quasi-equilibrium such as mean force calculation
or umbrella sampling are questionable for flexible polymers such as
single stranded DNA (ssDNA), which has a persistence length of about 3 nm, that is comparable to or smaller than the thickness of usual pores.
These difficulties are still not solved and form a very active field of research.

The electrostatic barrier was calculated in Ref. \cite{zhang07X}  for ssDNA in a biological
$\alpha$-hemolysin pore, and a value of
 more than 10 $k_B T$ was found. In these calculations the barrier is governed
by the dielectric mismatch effect between the surrounding water and the membrane material that
is less polarizable. This leads to a repelling force that drives uncompensated charges 
out of the pore. As this effect is of electrostatic origin it is obvious
that it can be screened by adding salt to the system.
This is consistent with the experimental findings that the rate with which
DNA enters such a pore can be increased by more than one order of magnitude when 
increasing the salt concentration from 0.25 mmol/l to 0.5 mmol/l \cite{bonthuis06a}.
To our knowledge no theoretical estimates of that barrier
has been made for larger pores and double stranded DNA (dsDNA). 
dsDNA is rather stiff with a persistence length of $\sim$50 nm at physiological conditions. 
Hence the 
configurational entropy of not too long strands is small. In these cases
the energetic barrier becomes more important to the dynamics of the translocation
process and the entropic barrier which is large only for longer or more flexible
chains becomes a secondary effect.

Previous findings based on simulations that take into account the dielectric contrast 
indicated that a dielectric
contrast increases the tendency to neutralize charges in the pore \cite{rabin05X} and squeezes the
counterion cloud \cite{guo09X}. The simulation works on this effect were restricted to
generic pore models with small diameters not much larger than a nanometer and small simulation cells,
because the employed scheme to solve Poisson's equation in presence of dielectric
discontinuities was computationally very demanding. Additionally it is questionable if
the specific properties of the biological pores that were modelled and the 
properties of the solvent in these small pores should not taken into account
in a more precise way. Works with fully electrostatic water models
also focused on small pores (e.g.~\cite{aksimentiev04X,luan08X}) and do not allow one
to distinguish polarization effects. With the recently developed \ICC\ algorithm
it is however possible to extend the range of systems available for MD 
simulations to larger pores and lower salt concentrations including dielectric boundary
effects. This is because \ICC\ can use arbitrary Coulomb solvers including the efficient and
highly optimized Coulomb solvers typically used in MD, that almost
scale linearly with the number of investigated charges. 
We looked specifically at a pore 5 nm in diameter which is about the smallest size of synthetic 
nanopores that is typically used in translocation experiments. Salt concentrations 
of no more than 10 mmol/l can be investigated while still being able to simulate a 
sufficiently large bulk reservoir. With a stiff model of dsDNA we investigated the capture 
barrier for two DNA segments of length
3 nm an 10 nm.

\section{Methods\label{sec:methods}}

\subsection{The \ICC\ algorithm}
In order to present the \ICC\ algorithm we make the following assumptions:
An arbitrarily shaped object with permittivity $\varepsilon_2$ is embedded
in a dielectric continuum with permittivity $\varepsilon_1$, referred to as the
outer and the inner medium. The Poisson equation in CGS units reads:
\begin{equation}
  \nabla \left(\varepsilon \nabla \Phi\right) = - 4 \pi \rho_\text{ext},
\end{equation} 
where $\Phi$ is the electrostatic potential, $\rho_\text{ext}$ a charge density that
represents e.g.~charged objects, and $\varepsilon$ a generic, position-dependent,
dielectric permittivity.
Solving this equation in direct proximity of the boundary leads to the well known
discontinuity of the normal component of the electric field
\begin{eqnarray}
  \varepsilon_1 \vec{E}_1 \cdot \vec{n} = \varepsilon_2 \vec{E}_2 \cdot \vec{n}.
  \label{eqn_jump}
\end{eqnarray}
This dielectric jump can be reproduced locally by placing an infinitesimal charged disk 
in the outer medium, parallel to the boundary, if the charge density of the disk fulfills:
\begin{eqnarray}
  \varepsilon_1 \left( \vec{E} - \frac{2\pi \sigma}{\varepsilon_1} \vec{n}  \right)\cdot  \vec{n} =
  \varepsilon_2 \left( \vec{E} + \frac{2\pi\sigma}{\varepsilon_1}  \vec{n} \right)\cdot  \vec{n} .
  \label{eqn_sigma1}
\end{eqnarray}
Here $\vec{E}$ denotes the field as measured in the outer medium if
the charged disk was not present. If the boundary is successively
replaced by these charged surface segments the emerging field is
identical to that of the original system. The \ICC\ algorithm is an
iterative procedure that allows to obtain a discretized analogous of
this surface charge on a grid of arbitrary shape. Any Coulomb solver
can be used to obtain the value of $\vec{E}$ at each position of the
discretized boundary.  Coulomb solvers employed in molecular dynamics
simulations can only handle point charges rather than charged
disks. The far field of a charged disk, however, is indistinguishable
from that of a point charge, so that at a certain distance from the
surface (one or two disk diameters) the proper solution is recovered
within good accuracy.  In close proximity to the disk this
approximation is not valid anymore; this implies two restrictions: On
the one hand, charged objects in a molecular dynamics simulations must
not approach closer that the disk radius (which is equivalent to
distance between adjacent surface elements). On the other hand the
mutual influence of adjacent surface segments carries the same
inaccuracy in the near-field. For a plane boundary this inaccuracy is
of no importance, since only the normal component of the electric
field determines the charge at the boundary, which then vanishes. In
presence of curved boundaries however the normal component is nonzero,
therefore the discretization has to be refined so that the angle
between the normal vectors of adjacent discretization points is small.
We have found that employing a discretization point distance that is
smaller than half of the local radius of curvature is sufficient.

The \ICC\ algorithm works as follows: Initially, a surface discretization with
appropriate normal vectors $\vec{n}_i$ and the corresponding surface element size
$A_i$ is chosen. Each of the surface segments is initialised with
a small but nonzero value to probe the local field. With a fast Coulomb solver the
field $\vec{E}_i$ at each of the discretization points is calculated. Then a new
charge is assigned to each surface point following the scheme:
\begin{eqnarray}
  q_\text{new} = \left(1-\lambda\right) q_\text{old} + \lambda A_i \sigma_i
\end{eqnarray}
where $\sigma_i$ is obtained from eq.~\ref{eqn_sigma1}:
\begin{eqnarray}
  \sigma_i = \frac{\varepsilon_1}{2 \pi} \frac{\varepsilon_1 - \varepsilon_2}{\varepsilon_1 + \varepsilon_2} \vec{E_i}\cdot\vec{n_i}.
\end{eqnarray}
This procedure is iterated until a self-consistent solution is obtained.
The factor $\lambda$ is a free parameter that determines the stability and the speed
of convergence of the relaxation scheme. It has turned out that a value of 0.9 
yields perfect stability for all tested cases and optimal speed of convergence. 
This iteration scheme does not need to be repeated after every MD step, but only after
the particle positions have changed noticeably, typically after 5-50 MD steps. 
Additional the small change of particle positions leads only to a small change in the boundary element
charge, so each \ICC\ update usually only needs 1-2 iteration steps.

\subsection{The DNA model and simulation details}
For the simulation we used the following model and parameters: The dsDNA molecule
was represented as a chain of spherical beads (50 beads per nm) forming a cylinder, constrained to fixed
positions.
Equal charges were assigned to each bead, in order to reach the line charge density of dsDNA
(\mbox{$2e/0.34$ nm}). Mobile counterions, as well as salt ions,  
were represented as monovalent point
charges, mutually interacting -- besides Coulomb interaction -- with the purely
repulsive Weeks--Chandler--Anderson (WCA) potential
\begin{equation}
U_\textsc{lj}(r)= 
\begin{cases}4 \epsilon_{\textsc{lj}} \left(\frac{z}{\sigma}\right)^{12} -
\left(\frac{r}{\sigma}\right)^{6} + \epsilon_\textsc{lj}   & \text{if }r < 2^{1/6}\sigma \\
0 &\text{otherwise,}
\end{cases}
\end{equation}
where $\sigma=0.425$ nm corresponds to an average size of the ions, including the first
hydration shell, 
and $\epsilon_\textsc{lj}=k_BT$. DNA beads and mobile
ions interact also via a WCA potential to mimic the steric repulsion of dsDNA, whereas
in this case \mbox{$\sigma=2.225$ nm}, corresponding to a DNA diameter of 2 nm (the value of
$\epsilon_\textsc{lj}$ being unaltered).  The temperature was fixed to \mbox{300 K},
and the Bjerrum length to \mbox{0.7 nm}. The pore was modeled as a reflecting
structureless, cylindrical cavity in a \mbox{8 nm} thick wall. The pore diameter was
set to \mbox{5 nm}, and the pore openings were smoothed with a torus shape
of radius 1 nm. \\

The system composed of the DNA fragment, mobile ions and the pore was set up in a cubic
simulation box with edge lengths of \mbox{20 nm}, with periodic boundary conditions in all three
directions. The electrostatic interaction
was calculated using the P$^3$M algorithm\cite{hockney88X}, choosing the parameters so to minimize the 
errors in the calculated forces \cite{deserno98X, deserno98bX, arnold05X}.
The time-step in simulation units was set to 0.002, and the \ICC\ algorithm was applied
every 10 integration steps. This assured that at least 99.99\% of the particle moved less then
\mbox{0.1 nm} between \ICC\ updates. The simulations were performed in the canonical
ensemble by integrating the Langevin equation using the velocity Verlet scheme.\\

The calculations were performed with different salt concentrations between 0 and 10 mmol/l
and DNA segments of two different lengths: a small fragment of 3 nm corresponding to about
15 base pairs and a longer segment of 10 nm corresponding to about 40 base pairs.  These
salt concentrations cover the range from and idealized salt free case to 10 mmol/l.
All simulations were performed with the freely available simulation package
ESPResSo \cite{limbach06}.

\subsection{Free Energy Profile}
The main subject of this investigation is the free energy profile of the model rigid DNA
fragment across a nanopore as generated by the rearrangement of the mobile
ion distribution and by the influence of the dielectric mismatch.  As we want to consider
the electrostatic aspects of the barrier of a stiff molecule, it is reasonable to
constrain the molecule to the path where we expect the lowest electrostatic barrier:
the motion in the pore axis, where the nearest approach to the pore wall is minimal.
This lowers the dimensionality of the problem and thus decreases computational complexity.
To get the free energy profile as a function of the only reaction coordinate $Z$,
the axial coordinate of the DNA fragment center the following simple a thermodynamic
integration is used. 

Recovering the free energy profile, as the potential of mean force is then, for this system,
straightforward\cite{ciccotti05a, kirkwood35a}. Only one DNA degree of freedom is left 
-- the position along the pore axis, $Z$, 
-- and it is employed as a reaction coordinate describing the pore crossing.
The derivative of the free energy with respect to the DNA fragment center $s$ 
can be therefore be written as
\begin{eqnarray}
  \frac{\partial F}{\partial s} =  
  \frac{ \int \d \Gamma \frac{\partial U}{\partial Z} \exp(-\beta U)\delta(s-Z)} 
  { \int \d \Gamma  \exp(-\beta U) \delta(s-Z)}, 
\end{eqnarray}
or, in other words,
$$
\frac{\partial F}{\partial s} = \left\langle \frac{\partial U}{\partial Z} \right\rangle_{Z=s}.
$$
Here the symbol $\left\langle\ldots\right\rangle_{Z=s}$ identifies an ensemble where the
coordinate $Z$ is constrained to have the value $s$. Note that since both the reaction
coordinate and the constraints are linear in the Cartesian coordinates, there is no difference
between constrained or conditional probabilities, and the 
Fixman \cite{fixman78a} potential for this problem is
just a constant. The ``standard'' free energy therefore coincides with the geometrical 
one \cite{hartmann07X}, and no ambiguity is left about  which quantity is being computed.

The generalized
force coincides with the component of the (Cartesian) force  $\partial V / \partial s$ on the
DNA along the pore axis, where $V$ is the potential energy.  
Eventually, the free energy profile can be obtained by thermodynamic
integration as
\begin{eqnarray}
  F(s) = \int_{s_0}^s \d s \left\langle \frac{\partial U}{\partial Z} \right\rangle_{Z=s},
\end{eqnarray}
where $s_0$ defines the zero point of the free energy. We chose $s_0$ where the DNA is midway
from the two periodic images of the pore, which the best approximation possible of a DNA
molecule far away from the pore.  In the coordinate system used in this work this corresponds
to $z = 0$.  Finally we employed the simple rectangle method  
in order to perform the thermodynamic
integration, introducing systematic errors of order $\frac{ \d^2 F}{\d s^2}$, which were found
to be small compared to the statistical ones.

\section{Results and Discussion\label{sec:results}}
In order to determine the effect of the dielectric boundary forces we
performed simulations where the \icc\ was turned on reducing the
permittivity of the membrane material to $\varepsilon = 2$, and
simulations where the \icc\ was not active which corresponds to a
permittivity of $\varepsilon = 80$ in the whole simulation box.
The mean force was taken at equidistant points every 1 nm. \\

\begin{figure}[h]
\includegraphics[width=\columnwidth]{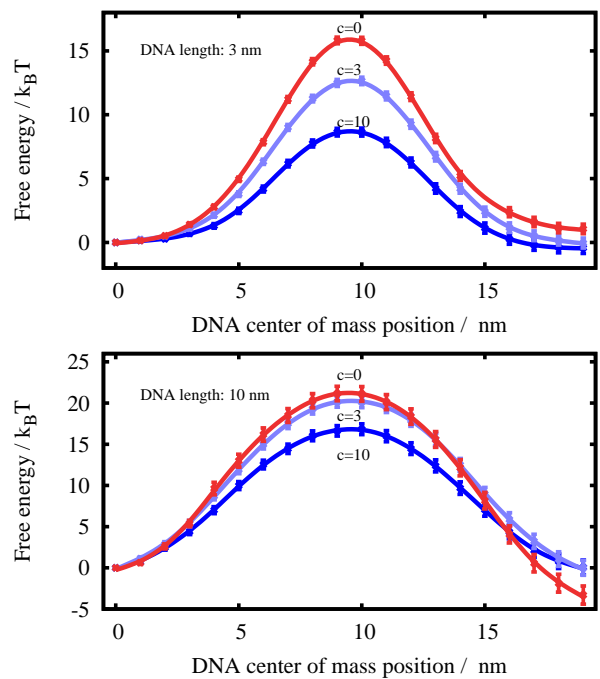}
\caption{\label{fig:pomf} Potential of mean force for two DNA-like
  polyelectrolytes of length 3nm (top) and 10 nm (bottom) in a cylindrical pore
  with $d = 5$ nm and $l =8$ nm taking into account dielectric boundary forces. 
  Increasing the salt concentration $c$ (given in mmol/l) reduces the effect of the dielectric boundary force.
  The apparent asymmetry of the graphs is statistically not significant.
  }
\end{figure}
\begin{table}
  \centering
  \begin{tabular}{| c | c | c | c | c |}
    \hline
    DNA len. & conc.& no.~counterions & no.~coions & barrier in $k_B T$ \\
    \hline
    3 nm& 0 mol/l & 17 & 0   & 15.7 $\pm$ 0.3 \\
        & 3 mol/l & 25 & 8   & 12.6 $\pm$ 0.3 \\
        & 10 mol/l& 46 & 29  & 8.6 $\pm$ 0.3 \\
    \hline
    10 nm & 0 mol/l & 58  & 0 & 21.1 $\pm$ 0.9 \\
          & 3 mol/l & 66  & 8 & 20.1 $\pm$ 0.7 \\
          & 10 mol/l& 87  & 29 & 16.8 $\pm$ 0.7 \\
    \hline
  \end{tabular}
  \caption{The number of ions used in the simulations and the determined free energy barriers for the 
   calculations taking into account the effect of dielectric boundary forces.}
  \label{tab:the_table}
\end{table}
Fig.~\ref{fig:pomf} shows the potential of mean force obtained for all
three salt concentrations taking into account the dielectric boundary
conditions for both DNA lengths. Table \ref{tab:the_table} summarizes the
barrier heights and the number of ions involved in each
simulation. The potential of mean force forms a bell-shaped barrier
whose height depends on the salt concentration. The height of the
barrier is striking as we obtain for the salt free case at least 15
$k_B T$ or 40 kJ/mol for the DNA segments of both sizes.  A barrier of
this size would slow down any process by six orders of magnitude.
However the barrier height decreases substantially with increasing salt
concentration.  For the short DNA strand the barrier is cut to half
the original height by adding 10 mmol/l monovalent salt.  For the
longer DNA strand the decrease of the barrier is smaller.  Surprising
is also the fact that the potential of mean force of the short DNA
strand is already more than 1 $k_B T$ at $z = $ 2 nm, where DNA end is
about 4 nm, so six Bjerrum lengths
away from the pore opening.  \\

For simulations without \icc\ we observed barriers of 1.5 $k_B T$ and 4 $k_B T$ for
the short and the long DNA strand independent on the salt concentration (not shown). 
In order to understand this barrier and 
to gain insight about the role of salt ions in the \icc\ case we turn to structural properties
of the ion distribution when the DNA is centered in the pore.

\begin{figure}[h]
\includegraphics[width=\columnwidth]{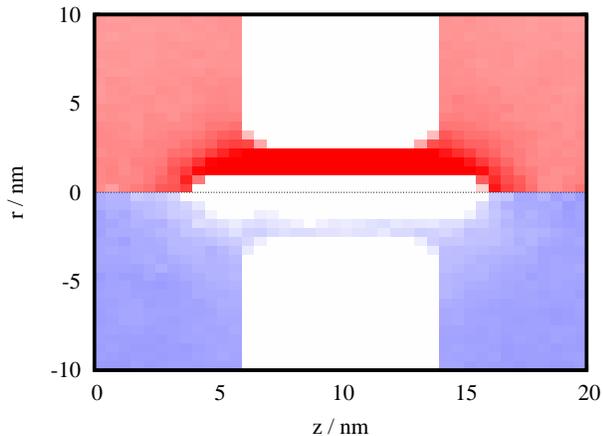}
\caption{\label{fig:density} The ion density as a function of the position $z$ along the pore axis and of the distance $r$ from it. A 10 nm long DNA 
fragment is centered in the pore. The upper part (red) reports the density of counterions and the lower part (blue) reports the density of 
coions. Darker colors indicate higher density. The increase in local
density of counterions and the depletion of coions close to the DNA
can easily be seen.}
\end{figure}

\begin{figure}[h]
\includegraphics[width=\columnwidth]{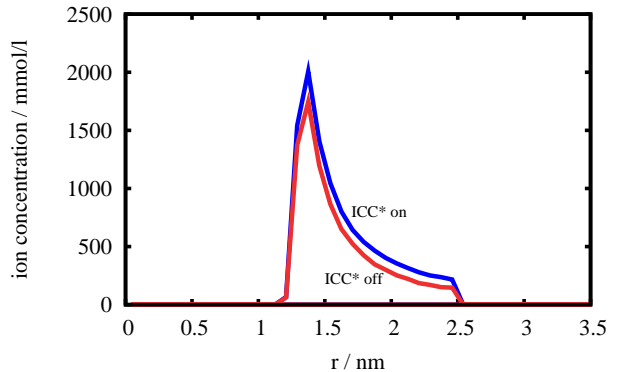}
\caption{\label{fig:rdf} The ion concentration in the pore as a function of the radius. 
In direct vicinity of the DNA a counter ion cloud is established. Applying \ICC\ 
increases the density of the screening cloud. The curves for the 10 mmol/l case and
the salt-free case are virtually identical so only one curve is shown.
}
\end{figure}

Fig.~\ref{fig:density} shows the ion density for the 10 millimolar case 
as a function of the radius and the axial
coordinate. The counterion density is strongly increased in direct proximity of the DNA.
Coions are depleted in the vicinity of the like-charged DNA molecule.
An exact investigation of the numbers shows however that far away from the DNA molecule the density
of anions and cations is identical. 
By looking at the radial distribution of
coions and counterions in the middle of the pore (fig.~\ref{fig:rdf}), it can be 
seen that the condensed ions exhibit a peak concentration of more than 1.5 mol/l, also for
the salt free case. Taking into account the dielectric boundary forces this effect is increased
as charge neutralization is amplified. This is similar to the effect that Rabin
and Tanaka \cite{rabin05X} found for smaller pores.

We also observe that the density of counterions does not decay to the bulk value and in fact
is higher than 200 mmol/l at the wall both for the
salt and the salt-free case. This shows that the pore confinement does not allow the formation
of a complete Debye layer, the high charge is thus only partly screened. This
leads to a compression of the counterion cloud in vicinity of the pore and to an effective 
entropic barrier, as observed also in the simulations without \ICC.




\section{Conclusions\label{sec:conclusions}}
In this work we focused on the effect of polarization
charges induced by the presence of a dielectric mismatch between the pore and the aqueous
environment in which the DNA molecule, its counterions, and salt ions
reside. In order to achieve this goal,  we employed a coarse-grained model that represents the DNA fragment with
overlapping charged beads. The mobile counterions and salt ions were modeled explicitly
and the induced surface charges were computed using the recently developed \ICC\ algorithm.

We computed the potential of mean force across the pore for two
polymer lengths (3 and 10 nm) at ionic strength ranging from zero to
10 mmol/l. The polarization contribution to the free energy barrier in
the salt-free case is remarkably high. The barrier was found to be
about 25 $k_B T$ for the longest fragment, whereas it is only 4 $k_B T$ 
in absence of the dielectric boundary force.  Polarization effects
have been shown to decrease significantly when only a small amount of
salt is added to the system.  In our model chain fluctuations and
deviations from an axial orientation of the DNA molecule have been
neglected. However, it is likely that also flexible molecules
experience similar barriers. This would strongly affect the kinetics
of the translocation process, but this effect can be decreased by
increasing the ionic strength of the surrounding solution.
\section{Acknowledgements}
We gratefully acknowledge financial support  by the DFG through SFB716-TP C5.  
We furthermore thank O.~A.~Hickey for a critical reading of the manuscript and the ESPResSo team for providing support.

\end{document}